\documentclass{emulateapj}

\slugcomment{Accepted in ApJ}

\shorttitle{Detection of SZ Decrement from SDSS LRGs in ACT Data}
\shortauthors{Hand et al.}

\usepackage{graphicx}
\usepackage[large]{subfigure}
\usepackage{amssymb, amsmath}
\usepackage[amssymb]{SIunits}

\def \bs #1{\boldsymbol{#1}}

\begin{document}

\title{The Atacama Cosmology Telescope: Detection of Sunyaev-Zel'dovich Decrement in Groups and Clusters Associated with Luminous Red Galaxies}

\author{
Nick~Hand\altaffilmark{1},
John~W.~Appel\altaffilmark{2},
Nick~Battaglia\altaffilmark{3},
J~Richard~Bond\altaffilmark{3},
Sudeep~Das\altaffilmark{4,2,1},
Mark~J.~Devlin\altaffilmark{5},
Joanna~Dunkley\altaffilmark{6,2,1},
Rolando~D\"{u}nner\altaffilmark{7},
Thomas~Essinger-Hileman\altaffilmark{2},
Joseph~W.~Fowler\altaffilmark{8,2},
Amir~Hajian\altaffilmark{3,1,2},
Mark~Halpern\altaffilmark{9},
Matthew~Hasselfield\altaffilmark{9},
Matt~Hilton\altaffilmark{10,11},
Adam~D.~Hincks\altaffilmark{2},
Ren\'ee~Hlozek\altaffilmark{6},
John~P.~Hughes\altaffilmark{12},
Kent~D.~Irwin\altaffilmark{8},
Jeff~Klein\altaffilmark{5},
Arthur~Kosowsky\altaffilmark{13},
Yen-Ting~Lin\altaffilmark{14,1,7,15},
Tobias~A.~Marriage\altaffilmark{1,16},
Danica~Marsden\altaffilmark{5},
Mike~McLaren\altaffilmark{5},
Felipe~Menanteau\altaffilmark{12},
Kavilan~Moodley\altaffilmark{10},
Michael~D.~Niemack\altaffilmark{8,2},
Michael~R.~Nolta\altaffilmark{3},
Lyman~A.~Page\altaffilmark{2},
Lucas~Parker\altaffilmark{2},
Bruce~Partridge\altaffilmark{17},
Reed~Plimpton\altaffilmark{5},
Erik~D.~Reese\altaffilmark{5},
Felipe~Rojas\altaffilmark{7},
Neelima~Sehgal\altaffilmark{18},
Blake~D.~Sherwin\altaffilmark{2},
Jonathan~L.~Sievers\altaffilmark{3},
David~N.~Spergel\altaffilmark{1},
Suzanne~T.~Staggs\altaffilmark{2},
Daniel~S.~Swetz\altaffilmark{5,8},
Eric~R.~Switzer\altaffilmark{19,2},
Robert~Thornton\altaffilmark{5,20},
Hy~Trac\altaffilmark{21},
Katerina~Visnjic\altaffilmark{2},
Ed~Wollack\altaffilmark{22}
}
\altaffiltext{1}{Department of Astrophysical Sciences, Peyton Hall, 
Princeton University, Princeton, NJ USA 08544}
\altaffiltext{2}{Joseph Henry Laboratories of Physics, Jadwin Hall,
Princeton University, Princeton, NJ, USA 08544}
\altaffiltext{3}{Canadian Institute for Theoretical Astrophysics, University of
Toronto, Toronto, ON, Canada M5S 3H8}
\altaffiltext{4}{Berkeley Center for Cosmological Physics, LBL and
Department of Physics, University of California, Berkeley, CA, USA 94720}
\altaffiltext{5}{Department of Physics and Astronomy, University of
Pennsylvania, 209 South 33rd Street, Philadelphia, PA, USA 19104}
\altaffiltext{6}{Department of Astrophysics, Oxford University, Oxford, 
UK OX1 3RH}
\altaffiltext{7}{Departamento de Astronom{\'{i}}a y Astrof{\'{i}}sica, 
Facultad de F{\'{i}}sica, Pontific\'{i}a Universidad Cat\'{o}lica,
Casilla 306, Santiago 22, Chile}
\altaffiltext{8}{NIST Quantum Devices Group, 325
Broadway Mailcode 817.03, Boulder, CO, USA 80305}
\altaffiltext{9}{Department of Physics and Astronomy, University of
British Columbia, Vancouver, BC, Canada V6T 1Z4}
\altaffiltext{10}{Astrophysics and Cosmology Research Unit, School of
Mathematical Sciences, University of KwaZulu-Natal, Durban, 4041,
South Africa}
\altaffiltext{11}{School of Physics \& Astronomy, University of Nottingham, NG7 2RD, UK}
\altaffiltext{12}{Department of Physics and Astronomy, Rutgers, 
The State University of New Jersey, Piscataway, NJ USA 08854-8019}
\altaffiltext{13}{Department of Physics and Astronomy, University of Pittsburgh, 
Pittsburgh, PA, USA 15260}
\altaffiltext{14}{Institute for the Physics and Mathematics of the Universe, 
The University of Tokyo, Kashiwa, Chiba 277-8568, Japan}
\altaffiltext{15}{Institute of Astronomy and Astrophysics, Academia Sinica, Taipei, Taiwan}
\altaffiltext{16}{Current address: Dept. of Physics and Astronomy, The Johns Hopkins University, 3400 N. Charles St., Baltimore, MD 21218-2686}
\altaffiltext{17}{Department of Physics and Astronomy, Haverford College,
Haverford, PA, USA 19041}
\altaffiltext{18}{Kavli Institute for Particle Astrophysics and Cosmology, Stanford
University, Stanford, CA, USA 94305-4085}
\altaffiltext{19}{Kavli Institute for Cosmological Physics, 
5620 South Ellis Ave., Chicago, IL, USA 60637}
\altaffiltext{20}{Department of Physics , West Chester University 
of Pennsylvania, West Chester, PA, USA 19383}
\altaffiltext{21}{Department of Physics, Carnegie Mellon University, Pittsburgh, PA 15213}
\altaffiltext{22}{Code 553/665, NASA/Goddard Space Flight Center,
Greenbelt, MD, USA 20771}

\begin{abstract}
We present a detection of the Sunyaev-Zel'dovich (SZ) decrement associated with the Luminous Red Galaxy (LRG) sample of the Sloan Digital Sky 
Survey. The SZ data come from 148 GHz maps of the equatorial region made by the Atacama Cosmology Telescope (ACT). The LRG sample is divided 
by luminosity into four bins, and estimates for the central SZ temperature decrement are calculated through a stacking process.  We
 detect and account for a bias of the SZ signal due to weak radio sources. We use numerical simulations to relate the observed decrement to  $Y_{200}$ 
 and clustering properties to relate the galaxy luminosity to halo mass. We also use a relation between brightest cluster galaxy luminosity and cluster mass based on stacked 
gravitational lensing measurements to estimate the characteristic halo masses. The masses are found to be around $10^{14} M_{\odot}$.
\end{abstract}

 
\section{Introduction}
New high resolution measurements of the sky at millimeter wavelengths are enabling the use of the thermal Sunyaev-Zel'dovich (SZ) effect as a precise cosmological 
tool. The effect arises from the inverse Compton scattering of the cosmic microwave background (CMB) and high energy electrons, usually occurring in the hot
 intra-cluster medium (ICM) of galaxy clusters \citep{sunyaev/zeldovich:1969, sunyaev/zeldovich:1970}. For frequencies below 218 GHz, the distortion manifests itself 
 as an arcminute-scale temperature decrement along the line of sight to a cluster. The amplitude of this decrement is nearly independent of redshift, making the SZ 
 effect particularly useful for examining the high redshift universe. The effect can also be used to explore the role of baryonic physics in cluster evolution, as its 
 amplitude is proportional to the thermal pressure of the ICM integrated along the line of sight. For a cluster in hydrostatic equilibrium, it is expected that the integrated 
 SZ signal scales with total cluster mass. For a full description of the SZ effect,  see the review articles by \citet{rephaeli:1995}, \citet{birkinshaw:1999}, and 
 \citet{carlstrom/etal:2002}.

Analyses of large X-ray and optically selected cluster samples have recently illustrated the cosmological potential of cluster surveys \citep[e.g.,][]{vikhlinin/etal:2009b, 
mantz/etal:2010, rozo/etal:2010}. The emergence  of large-area SZ cluster surveys, such as the South Pole Telescope \citep[SPT;][]{carlstrom/etal:2009} and the 
Atacama Cosmology Telescope  \citep[ACT;][]{fowler/etal:2007, swetz/etal:2010}, will supplement this previous work. Furthermore, upcoming results from the 
{\em{Planck}} satellite are expected to include a large number of SZ-detected clusters \citep{bartlett/etal:2008}. The SPT collaboration reported its first SZ detections in 
\citet{staniszewski/etal:2009} and has since identified twenty-two cluster candidates in \citet{vanderlinde/etal:2010}. Twenty-one of these clusters were optically 
confirmed in \citet{high/etal:2010}. The X-ray properties of the SPT sample and the derived SZ/X-ray relations are presented in a follow-up study 
\citep{andersson/etal:2010}. The ACT collaboration presented its first cluster detections in \citet{hincks/etal:2010} and has since reported on twenty-three clusters in 
\citet{marriage/etal:2010b}, all of which have been optically confirmed \citep{menanteau/etal:2010}. The cosmological implications of the ACT sample are discussed 
in \citet{sehgal/etal:2010}. 

In order to fully use cluster surveys for cosmological purposes, independent and robust estimates of the cluster masses are needed. Several studies have explored the 
relation between the integrated SZ signal $Y$ and total cluster mass $M$ for massive clusters ($M > 10^{14} M_{\odot}$) 
\citep[i.e.,][]{benson/etal:2004, bonamente/etal:2008, melin/etal:2010, huang/etal:2010, plagge/etal:2010}. The integrated SZ signal is defined as 
\begin{equation}
\label{Y_def}
Y = \int_{\Omega} y(\hat n) d\Omega,
\end{equation}
where $y$ is the usual Compton $y$-parameter and $\Omega$ is the solid angle of the cluster.  Assuming that thermal energy results solely from gravitational collapse,
 it is possible to derive self-similar scalings between observables (e.g., $Y$) and cluster mass. Furthermore, observations show that the SZ-mass relation is relatively 
 insensitive to the specifics of cluster physics (i.e., cooling, AGN feedback) \citep{bonamente/etal:2008, sehgal/etal:2010}.  

In this work, we report on the stacking of a subset of the ACT 2009 equatorial data at the positions of Luminous Red Galaxies (LRGs) measured by the Sloan Digitial Sky Survey
 (SDSS). Various studies have examined the clustering properties of the SDSS LRGs \citep[i.e.,][]{zheng/etal:2009, reid/spergel:2009} and have found that most of these galaxies
  reside in halos of typical mass $M \sim 10^{13}-10^{14} h^{-1} M_{\odot}$. This paper describes the stacking process and the detection of SZ signal after binning the sample by
   luminosity. Estimates for the characteristic halo mass of each luminosity bin are obtained using two methods: gravitational lensing mass measurements and  an empirical model for 
   halo bias as a function of luminosity \citep{tegmark/etal:2004, reid/etal:2010, reyes/etal:2008, zehavi/etal:2005} that describes the LRG sample well, as shown in \citet{percival/etal:2008}.  Finally, 
   we present an initial analysis of  the relation between integrated SZ signal and halo mass. Unless otherwise stated, we assume a flat $\Lambda$CDM cosmology with 
   $\Omega_m$ = 0.264,  $\Omega_{\Lambda}$ = 0.736, $\sigma_8$ = 0.80 and $H_{\circ}$ = 100\emph{h} km $\mathrm{s^{-1} \ Mpc^{-1}} $ with \emph{h} = 0.71
    \citep{komatsu/etal:2009}. 

This paper is organized as follows. In Section \ref{data}, we describe the LRG sample and the ACT map. In Section \ref{filtering}, we describe the process of filtering, and in Section
 \ref{methods} we discuss our methods for binning and stacking the ACT data. We present the stacking results in Section \ref{results} and examine the SZ-mass scaling relation in 
 Section \ref{scaling_relation}. We conclude in Section \ref{conclusion}.

\begin{figure*}[t]
	\begin{center}
	\subfigure{\includegraphics[scale=0.75]{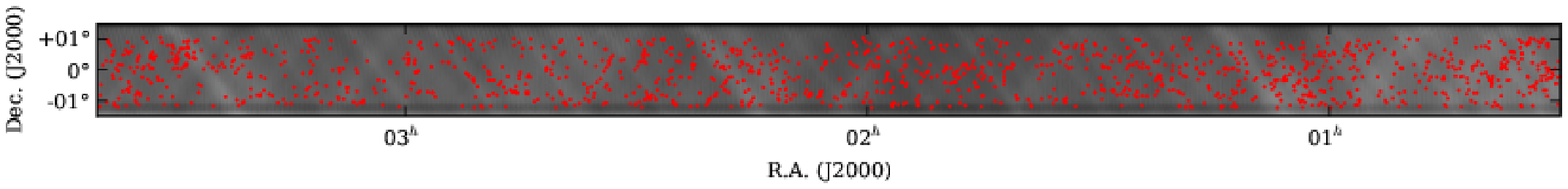}}
	\subfigure{\includegraphics[scale=0.75]{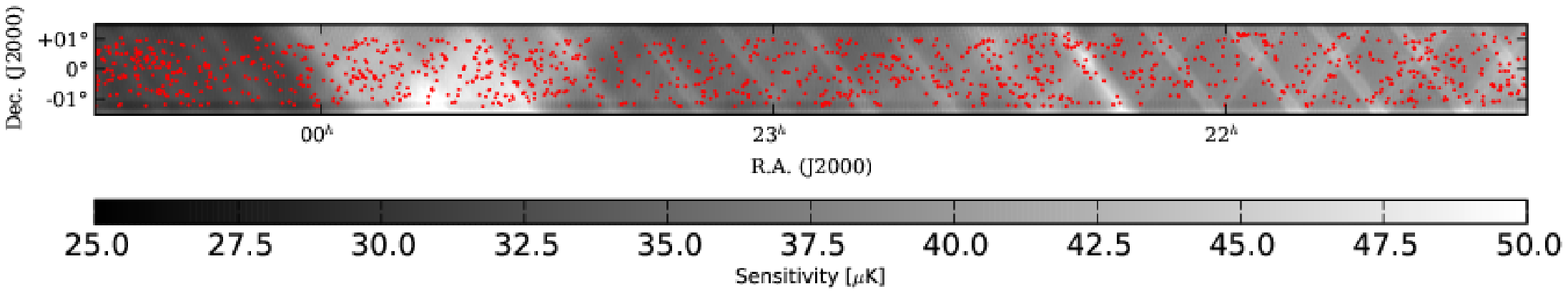}}
	\end{center}
\caption{\label{sensitivity} Sensitivity map with LRG locations. The map shows the sensitivity over the subset of the ACT 2009 148 GHz equatorial data considered for this study. 
The gray-scale encodes the noise rms in $\micro$K of a map match-filtered for cluster detection.  The median noise in the map is 34 $\micro$K in CMB temperature units.}
\end{figure*}

\section{Data}
\label{data}

In this section, we describe the subset of the ACT data used in this analysis (Sec. \ref{act_data}) and the LRG sample from the SDSS (\ref{lrg_data}). 

\subsection{ACT Data}
\label{act_data}

ACT is a six-meter telescope operating in the Atacama Desert of Chile at an altitude of 5200 meters. The telescope site was chosen for its dry climate as well as the ability to observe 
in both the northern and southern skies. The telescope has three 1024-element arrays of transition edge sensors, one each operating at 148 GHz, 218 GHz, and 277 GHz. For a more
 detailed introduction to the ACT instrument, observations, and data reduction, see \citet{fowler/etal:2007}, \citet{swetz/etal:2010} and \citet{das/etal:2010}. The science release from 
 the ACT 2008 survey of a region south of the Galactic equator includes results on the CMB power spectrum and related parameter constraints
  \citep{fowler/etal:2010, das/etal:2010, dunkley/etal:2010, hajian/etal:2010}. Results are also presented on compact millimeter sources \citep{marriage/etal:2010a} and clusters 
  \citep{marriage/etal:2010b, menanteau/etal:2010, sehgal/etal:2010}. The present study uses data at 148 GHz from a subregion of the 2009 equatorial survey. The subregion 
  covers 268 square-degrees and lies between right ascensions $21^{\mathrm{h}}20^{\mathrm{m}}$ and $03^{\mathrm{h}}40^{\mathrm{m}}$ and declinations $-01^{\circ}30'$ and 
  $01^{\circ}30'$. This region is coincident with the SDSS Stripe 82, which contains a rich multi-wavelength data set. Figure \ref{sensitivity} is a map of the sensitivity across the subregion
   of the study, along with the locations of the LRGs. The map has been match-filtered for cluster detection as described in Section \ref{filtering}. The median rms in the filtered map is 34 
   $\micro$K in CMB temperature units.

\subsection{LRG Sample}
\label{lrg_data}

The LRG sample used in this study is from the SDSS, a survey that has performed five-band ($ugriz$) photometry \citep{fukugita/etal:1996} using a specialized wide-field camera
 \citep{gunn/etal:1998} and multi-object spectroscopy using a pair of spectrographs. The survey has imaged one quarter of the sky at high Galactic latitude and conducted 
 spectroscopic follow up observations of approximately one million of the detected objects \citep{eisenstein/etal:2001, richards/etal:2002, strauss/etal:2002}. All data are processed 
 automatically by pipelines that detect and measure photometric properties of sources \citep{lupton/etal:2001}. The SDSS has had seven major data releases
  \citep{abazajian/etal:2003, abazajian/etal:2004, abazajian/etal:2005, adelman_mccarthy/etal:2006, adelman_mccarthy/etal:2007, adelman_mccarthy/etal:2008, abazajian/etal:2009}. 

The LRG catalog used in the current work is drawn from the sample prepared by \citet{kazin/etal:2010}, which is taken from the seventh data release  (DR7) of the SDSS. The data 
set was first developed by \citet{eisenstein/etal:2001} and is publicly
 available.\footnote[1]{\href{http://cosmo.nyu.edu/~eak306/SDSS-LRG.html}{http://cosmo.nyu.edu/$\sim$eak306/SDSS-LRG.html}} The DR7 LRG sample contains $\sim$110,000
  galaxies and extends to redshift $z \simeq 0.5$.  The subset used in this study contains 2681 LRGs that fall within the 268 square-degree ACT map. The subset has a spectroscopic 
  redshift range 0.16 $< z <$ 0.47 and has rest frame $g$-band absolute magnitudes $-23.2  < M_g <  -21.2$. K corrections \citep{blanton/roweis:2007} have been applied to the 
  galaxies. The SDSS catalog of LRGs serves as a good tracer of large scale structure, as LRGs are highly biased and trace more massive dark matter halos than other galaxy samples. 
  They also exhibit a distinctive 4000 \AA \ break in their spectral energy distributions, making redshift determination easier than for other galaxies. 

\begin{table*}[t]
\caption{Luminosity Bins and Halo Masses for Stacked LRGs}
\begin{center}
\label{mass_table}
\begin{tabular}{| c c c c c c  c |}
\hline
Bin & $N_{\mathrm{bin}}$  &  $ \langle L_{0.1r} \rangle$  & $L_{0.1r}$ Range   & $\langle z \rangle $  & $M_{200\bar{\rho}}$\footnotemark[1]  & $M_{200\bar{\rho}}$\footnotemark[2]   \\
       &                  & $10^{10}L_{\odot}$  &  $10^{10}L_{\odot}$  & & $10^{14} M_{\odot}$ & $10^{14} M_{\odot}$  \\  \hline
1 & 26    & 30.1 &  25.9 - 44.9   &  0.39 & 4.26 $\pm$ 1.06 & 2.59 $\pm$ 0.87  \\
2 & 60    & 23.6  & 21.8 - 25.3   &  0.40 & 2.89  $\pm$ 0.20  & 2.05 $\pm$ 0.43 \\
3 & 674  &  16.7   &  13.8 - 21.8 & 0.37 &  1.71  $\pm$ 0.34 & 1.43 $\pm$ 0.31 \\
4 & 1921 & 10.0  & 1.1 - 13.8  &  0.30 & 0.80 $\pm$  0.24  &  0.82 $\pm$ 0.22 \\
\hline
\end{tabular}
\footnotetext[1]{Determined from analysis of halo bias.}
\footnotetext[2]{Determined from lensing measurements in \protect\citet{reyes/etal:2008}.}
\end{center}
\end{table*}

\section{Filtering the ACT Data}
\label{filtering}

To maximize the SZ signal of the clusters associated with the LRGs, we employ a matched filter \citep{haehnelt/tegmark:1996, herranz/etal:2002, melin/etal:2006}. The filtering scheme 
used here closely follows \citet{marriage/etal:2010a}, with exceptions noted below. At a position $\bs{x}$ in the map, we model the temperature measured as the sum of the SZ signal
 plus the other components in the map (CMB fluctuations, atmospheric fluctuations, instrument noise and point sources):

\begin{equation}
\delta T (\bs{x}) = \sum_i \delta T_{\circ, i} b_{i}(\bs{x} - \bs{x_i}) + \delta  T_{\mathrm{other}}(\bs{x}),
\end{equation}
where $\delta T_{\circ,i}$ is the peak amplitude of the $i^{\mathrm{th}}$ LRG and $b_i$ is the unit-normalized ACT 148 GHz beam function, taken to be isotropic with a FWHM of 
$1.4'$ \citep{hincks/etal:2010}. We first mask the brightest sources in the map, which amount to 146 objects. Only pixels associated with a source with SNR $\ge$ 5 are masked. 
These are predominantly radio sources \citep{marriage/etal:2010a}. None of these sources lies within $0.5'$ of an LRG position. Before applying the matched filter, we multiply the
 map, pixel-wise, by the number of observations per pixel normalized by the maximum number of observations in any single pixel, $\sqrt{N_{\mathrm{obs}}(\bs{x}) / N_{\mathrm{obs, max}}}$. 
 This weighting accounts for local changes in the amplitude of the white noise and is equivalent to down-weighting pixels with a high white noise rms. This weight scheme is also 
 used during the stacking process when determining the matched filter decrement $\delta T_{\circ}$ (see Section \ref{stacking_algorithm}). Next, the map is filtered in Fourier space 
 using a matched filter: 

\begin{equation}
\label{filter}
\delta T_{\mathrm{filt}}(\bs{k})= \frac{\tilde{b}^{\star}(\bs{k})  \ | \tilde{\delta T}_{\mathrm{other}}(\bs{k}) |^{-2}  \ \tilde{\delta T}(\bs{k})}{\int \tilde{b}^{\star}(\bs{k'})
  \ | \tilde{\delta T}_{\mathrm{other}}(\bs{k'}) |^{-2} \ \tilde{b}(\bs{k'}) d\bs{k'} }.
\end{equation} 
Since the power spectra of the atmosphere and the CMB scale as $l^{-4}$ for $l > 1000$,  the filter is approximately a fourth derivative operator convolved with the beam. 
The choice of the beam function $b$ for the spatial profile of the cluster differs from the method used in \citet{marriage/etal:2010b}, which employs a $\beta$-model for the cluster 
profile. From comparisons with simulations, we expect the ACT 148 GHz beam function to be a good match for the spatial profile of clusters and hence adopt it in the matched filter. 
 It is particularly effective at reducing scatter from CMB noise, to which broader filters are more sensitive.  

Rather than explicitly modeling the noise contribution $\delta  T_{\mathrm{other}}(\bs{x})$, we use the original input map with sources masked for the power spectrum of
 $\tilde{\delta T}_{\mathrm{other}}(\bs{k})$ in equation \ref{filter}. As the model for $\delta  T_{\mathrm{other}}(\bs{x})$ is not well-conditioned, we smooth
  $| \tilde{\delta T}_{\mathrm{other}}(\bs{k}) |^{-2}$ with a Gaussian filter ($\sigma$ = 5 Fourier space pixels). We set outlying Fourier space pixels with amplitude greater than 10 times the
   median pixel value  to the median value. We also apply an additional low-$\ell$ filter in Fourier space that grows from zero at $\ell = 0$ to unity at $\ell = 1200$ as $\mathrm{sin}^5(\pi \ell / 2400).$ This filter accounts for any modification of CMB statistics due to multiplying by the non-uniform weight map and makes our result more robust to any potential non-Gaussianity in the large-scale noise that might arise from residual atmospheric noise in the maps.  After finishing the filtering process, we do not include any sources that fall within $10'$ of the map edge  in order to mitigate noise from edge effects. This reduces the map size from $284$ square degrees to 268 square-degrees. The entirety of the LRG sample discussed previously (2681 objects) falls within the bounds of the reduced map, leading to roughly 10 LRGs per square-degree.

\section{Stacking Methods}
\label{methods}
While ACT can easily detect massive clusters ($M > 10^{15} M_{\odot})$ \citep{marriage/etal:2010b}, CMB fluctuations and detector noise limit the direct detection of less 
massive clusters. Since the mean of both the CMB fluctuations and atmospheric noise should be zero, we stack submaps around LRGs to enable the detection of weaker SZ signals.

\subsection{Luminosity Bins}
\label{luminosity_bins}
Before stacking the LRGs, we bin the sample by $r$-band luminosity K corrected to $z = 0.10$, $L_{0.1r}$. Table \ref{mass_table} shows the four luminosity bins used in this analysis. 
This table also lists mass estimates for each bin, as described in Section \ref{mass_estimation}. The luminosities have been computed using the Galactic extinction-corrected 
\citep[using maps from][]{schlegel/etal:1998} Petrosian magnitudes from the DR7 CAS and their measured errors. They have been K corrected to $z = 0.10$ using the ${\tt kcorrect\_v4\_2}$
 software package \citep{blanton/roweis:2007}. There have been no added evolutionary corrections. We also refer to the K corrected $^{0.1}r$-band absolute magnitude of an LRG as
  $M_{0.1r}$. Following \citet{blanton/etal:2003},
\begin{equation}
M_{0.1r} = m_r - DM(z, \Omega_{m}, \Omega_{\Lambda}, h)  - K_{0.1 rr}(z),
\end{equation}
where DM($z, \Omega_{m}, \Omega_{\Lambda}, h$) is the distance modulus in a $(\Omega_{m}=0.3, \Omega_{\Lambda}=0.7, h=0.71)$ cosmology and $K_{0.1rr}(z)$ is the K correction 
from the $r$-band of a galaxy at redshift $z$ to the $^{0.1}r$-band. Luminosities are reported in solar luminosities, where we have used $M_{\odot, 0.1r}$ = 4.76 \citep{blanton/etal:2003}.
 These definitions of luminosity and magnitude will facilitate the estimation of halo masses for our binned data in Section \ref{mass_estimation}.

\subsection{Algorithm}
\label{stacking_algorithm}
We stack filtered submaps with area $10'$ x $10'$ centered on each LRG in a given luminosity bin, and then estimate the matched filter decrement $\delta T_{\circ}$ from the central 
temperature value in the stacked map. Recall that the filtered map is weighted by the number of observations per pixel normalized by the maximum number of observations in any single 
pixel. To minimize the errors induced by the $0.5'$ pixel size of the ACT maps \citep{marriage/etal:2010a}, the pixel size of the submap centered on each LRG is decreased using Fourier 
interpolation. For the present analysis, the pixel size is decreased to $0.03125'$ and the submaps are recentered.  Before determining $\delta T_{\circ}$, the stacked map for a given bin is 
convolved with the ACT 148 GHz beam function $b$ to recover decrement information from the surrounding pixels. Lastly, we sum over the five pixels within $4''$ of the centroid of the 
stacked map and divide by the sum of the weights to compute  $\delta T_{\circ}$. We compute the weight for each submap in the stack before decreasing the pixel size and use the value of 
$\sqrt{N_{\mathrm{obs}} / N_{\mathrm{obs, max}}}$ at the central $0.5'$ pixel.  

 \begin{table*}
\caption{Central SZ Temperature Decrement and Integrated SZ Signal for All Stacked LRGs, Binned by $L_{0.1r}$}
\begin{center}
\label{temp_full_table}
\begin{tabular}{| c c c c  c c |}
\hline
Bin & $N_{\mathrm{bin}}$ & $\delta T_{\circ}$ &  SNR & $Y_{200\bar{\rho}}$ & $Y_{200\bar{\rho}} \ d^2_A \ E(z)^{-2/3}$ \\
         &  & $\micro$K       &    & $\mathrm{arcmin^2}$ & $\mathrm{Mpc^2}$   \\ \hline
1 & 26 & $-27.5 \ \pm$ 11.3 & 2.4 & (2.0 $\pm$ 0.7) $\times 10^{-4}$ & (1.4 $\pm$ 0.6) $\times 10^{-5}$  \\
2 & 60 & $-9.9 \  \pm$ 5.3 & 1.9 & (6.2 $\pm$ 2.7) $\times 10^{-5}$ & (4.4 $\pm$ 2.4) $\times 10^{-6}$ \\
3 & 674  &$ -4.7 \  \pm$ 1.5 & 3.1 & (2.4 $\pm$ 0.7) $\times 10^{-5}$ & (1.7 $\pm$ 0.6) $\times 10^{-6}$ \\
4 & 1921 & $-1.6 \ \pm$ 1.0 & 1.6 & (4.5 $\pm$ 4.3) $\times 10^{-6}$ & (4.6 $\pm$ 2.8) $\times 10^{-7}$  \\
\hline
 \end{tabular} 
 \caption{Central SZ Temperature Decrement and Integrated SZ Signal for Stacked Radio-quiet LRGs, Binned by $L_{0.1r}$}
 \label{temp_rq_table}
\begin{tabular}{|c c c c c c  |}
\hline
Bin & $N_{\mathrm{bin}}$ & $\delta T_{\circ}$ & SNR  & $Y_{200\bar{\rho}}$ & $Y_{200\bar{\rho}} \ d^2_A \ E(z)^{-2/3}$ \\
         &  & $\micro$K   &  & $\mathrm{arcmin^2}$ &  $\mathrm{Mpc^2}$  \\ \hline
1 & 21 & $-28.2 \ \pm$ 12.9  & 2.2 &  (2.2 $\pm$ 0.8) $\times 10^{-4}$ & (1.5 $\pm$ 0.7) $\times 10^{-5}$  \\
2 & 51  &$-10.6 \ \pm$ 4.9 & 2.2 & (5.4 $\pm$ 2.4) $\times 10^{-5}$ & (4.7 $\pm$ 2.2) $\times 10^{-6}$ \\
3 &  587 &$ -6.1 \ \pm$ 1.6 &  3.8 & (3.3 $\pm$ 0.7) $\times 10^{-5}$ & (2.3 $\pm$ 0.6) $\times 10^{-6}$  \\
4 & 1732 & $-2.1 \ \pm$ 1.0 & 2.1 & (1.1 $\pm$ 0.9) $\times 10^{-5}$ & (7.9 $\pm$ 6.2) $\times 10^{-7}$    \\
\hline
 \end{tabular} 
 \tablecomments{The estimates for $Y_{200\bar{\rho}}$ are determined from $\delta T_{\circ}$ using microwave sky simulations \citep{sehgal/etal:2009}. They correspond to the halo
  masses $M_{200\bar{\rho}}$ derived from considering luminosity-dependent halo bias (Table \ref{mass_table}). The uncertainty in $\delta T_{\circ}$ is determined from bootstrap resampling. The central decrements and associated errors are measured in CMB temperature units.}
 \end{center}
 \end{table*}

\section{Stacking Results}
\label{results} 

\subsection{Matched Filter Decrement $\delta T_{\circ}$}
\label{decrement_results}
Tables  \ref{temp_full_table} and \ref{temp_rq_table} contain estimates for the matched filter decrement $\delta T_{\circ}$ obtained for each of the four luminosity bins. Due to radio
 contamination (see discussion below), we report results for the full LRG sample as well as a radio-quiet subset, which excludes approximately 10\% of the LRG sample. The value
  for $\delta T_{\circ}$ reported in Tables \ref{temp_full_table} and \ref{temp_rq_table} is the temperature decrement in the ACT map, recovered from the matched filter. Figure 
  \ref{stacked_images} shows $7' \times 7'$  stacked submaps for each luminosity bin for both LRG samples. Stacked maps obtained when using random LRG positions are also 
  shown, and it is apparent that there is no clear detection in these maps, as expected. 

In addition to the central peak in the stacked images,
several other significant peaks are evident, particularly
the deep cold spot at the upper edge of the stacked
map for Bin 2, which has a depth of around 24 $\micro$K and is significantly
deeper than the stacked cluster at the center. For this stacked map, the pixel noise is around 5 $\micro$K, making
the upper-edge cold spot a 5$\sigma$ discrepancy. Examination
of the 60 individual maps in the stack reveals that
two maps contain decrements of $<$ 150 $\micro$K at the location of the cold spot. In addition, 11 additional map sections show
decrements larger than 50 $\micro$K at the approximate position of
the upper-edge cold spot. This large number of decrements in similar
positions relative to the LRGs in our catalog clearly is not
a random phenomenon, but rather reflects the fact that galaxy clusters and groups
are strongly clustered on the sky.

\begin{figure*}[t]
	\begin{center}
	\subfigure[Full LRG Sample]{\includegraphics[scale=0.7]{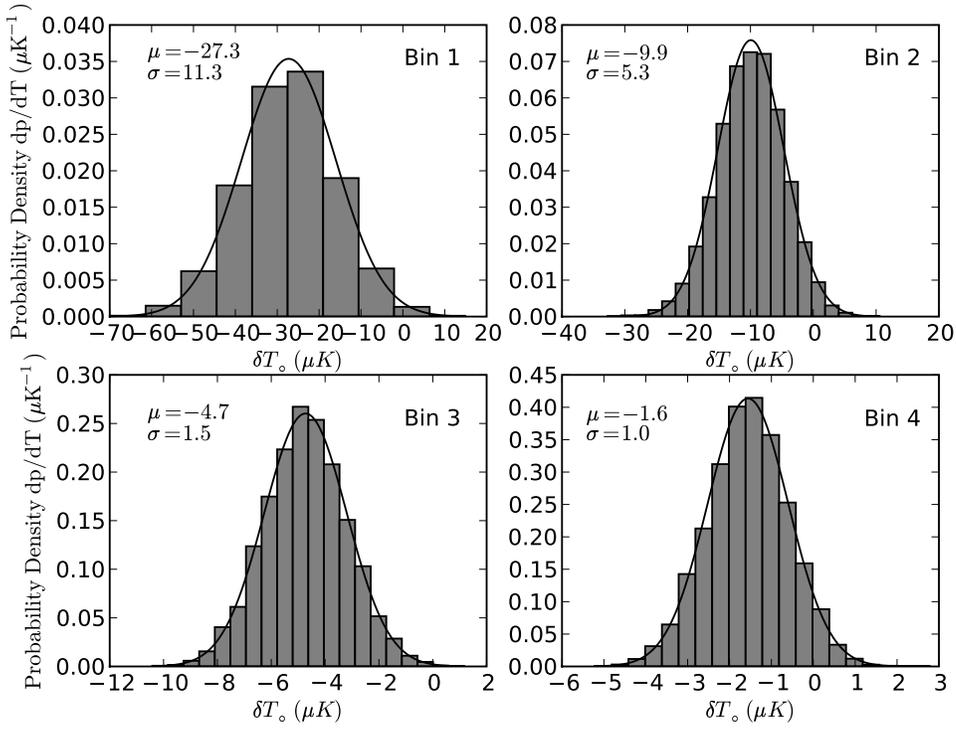}}
	\subfigure[Radio-quiet LRG Sample]{\includegraphics[scale=0.7]{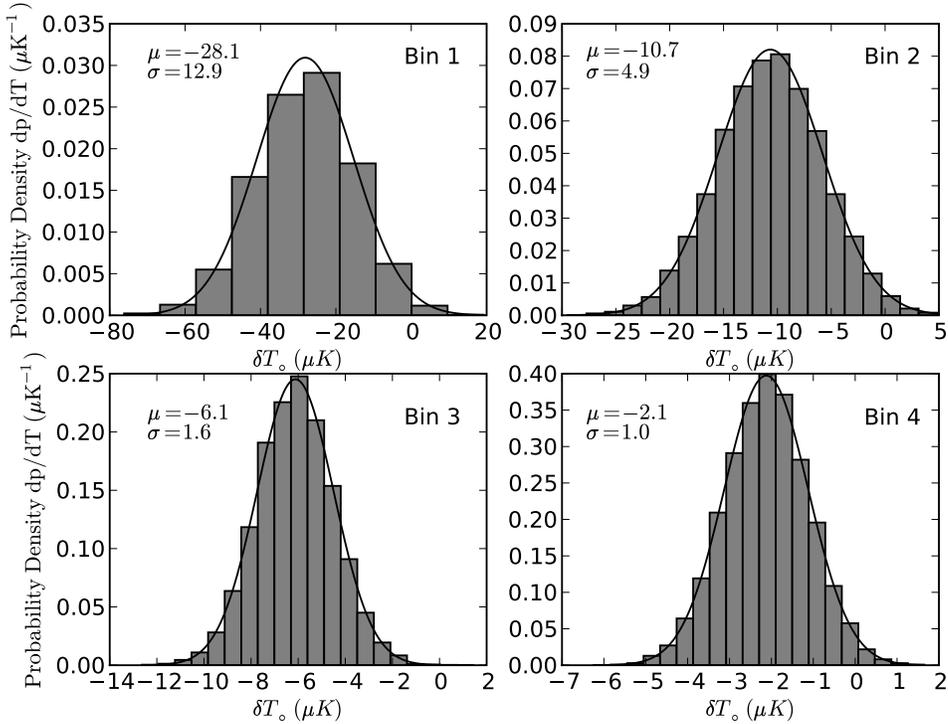}}
	\end{center}
\caption{\label{mean_distribution} The distribution of the mean of $\delta T_{\circ}$ for each luminosity bin in the full and radio-quiet LRG samples, determined from bootstrap resampling. 
The mean $\mu$ and standard deviation $\sigma$ obtained from the resampled distribution are shown for each bin. The distribution of the mean is well-approximated by a Gaussian function,
 which is shown for each panel. The bootstrap errors $\sigma$ are reported in Tables \ref{temp_full_table} and \ref{temp_rq_table} and are used throughout this work as the error 
 estimate for $\delta T_{\circ}$.}
\end{figure*}

The errors in $\delta T_{\circ}$ presented in Tables \ref{temp_full_table} and \ref{temp_rq_table} are computed for each bin using bootstrap resampling.\footnote[1]{We create 10,000 bootstrap realizations from the values of $\delta T_{\circ}$ in each bin and use the distribution of the resulting bin averages to estimate the uncertainty} These errors measure the scatter in
 the properties of individual LRGs. Figure \ref{mean_distribution} shows the resampled distribution of the mean of $\delta T_{\circ}$ for each of the four luminosity bins for the full and radio-quiet LRG samples. 
The distribution of each bin is well-approximated by a Gaussian function. 

\begin{table}[b]
\caption{Matched Filter Decrement $\delta T_{\circ}$ for Various Systematic Tests of the ACT Data}
\begin{center}
\label{tests_table}
\begin{tabular}{ |c c c |}
\hline
\hline
Test &  $N_{\mathrm{bin}}$ &  $\delta T_{\circ}$ $(\micro$K)  \\ \hline
Top & 1295  & $ -2.51 \ \pm$ 1.16 \\
Bottom & 1386 &$ -3.13 \ \pm$ 1.10 \\
Left  & 1406   & $-2.94 \ \pm$ 0.99 \\
Right & 1275 & $-2.67 \  \pm$ 1.35 \\
High $N_{\mathrm{obs}}$ & 1455 & $-2.15 \ \pm$  0.98\\
Low $N_{\mathrm{obs}}$ & 1226 & $-3.85 \ \pm$ 1.36  \\
\hline
Radio-loud& 290 & 3.27 $\pm$ 2.65  \\
Radio-quiet & 2391 & $-3.56 \ \pm$ 0.84 \\
\hline
\hline
 \end{tabular} 
 \tablecomments{The significant negative signal shown is expected. The central decrements and associated errors are measured in CMB temperature units.}
 \end{center}
 \end{table}
 
 Table \ref{tests_table} contains the matched filter decrements $\delta T_{\circ}$ obtained when stacking various subsets of the LRG sample. These subsets are designed to test various
 systematic aspects of the ACT data. The first four tests in Table \ref{tests_table} probe spatial discrepancies in the data by splitting the ACT map by right ascension and declination. We
  conclude that the signal is consistent throughout the map. We also examine variations in white noise amplitude by splitting the sample spatially by the number of observations
   $N_{\mathrm{obs}}$ per pixel. We determine the median value of $N_{\mathrm{obs}}$ per pixel and then stack subsets of LRGs that lie above and below this median value. We
    conclude that the signal is present for both high and low noise bins in the map. 

Radio source contamination at 148 GHz appears to have a significant effect.  Using the  SDSS DR7 CAS, we identified LRGs with radio counterparts within $2''$ in  the 1.4 GHz VLA 
FIRST Survey \citep{becker/etal:1995} and define these LRGs to be radio-loud. There are  290 radio-loud\ LRGs in the sample: 5/26 in Bin 1, 9/60 in Bin 2, 87/674 in Bin 3 and 189/1921 
in Bin 4. Contamination by FIRST radio sources is potentially a concern for $\sim$$10\%$ of the LRG sample. We also compute the matched filter decrement for each bin 
excluding the 290 radio-loud LRGs.  While the stacked radio-quiet LRGs show a significant ($> 4 \sigma$) decrement, the radio-loud LRGs show clear evidence of radio source 
contamination. For the rest of this analysis, we report results for both the full and radio-quiet samples (see Tables \ref{temp_full_table} and \ref{temp_rq_table}).

\subsection{Integrated SZ Signal Y} 

\begin{figure*}[t]
	\begin{center}
	\subfigure[Full LRG Sample]{\includegraphics[scale=0.75]{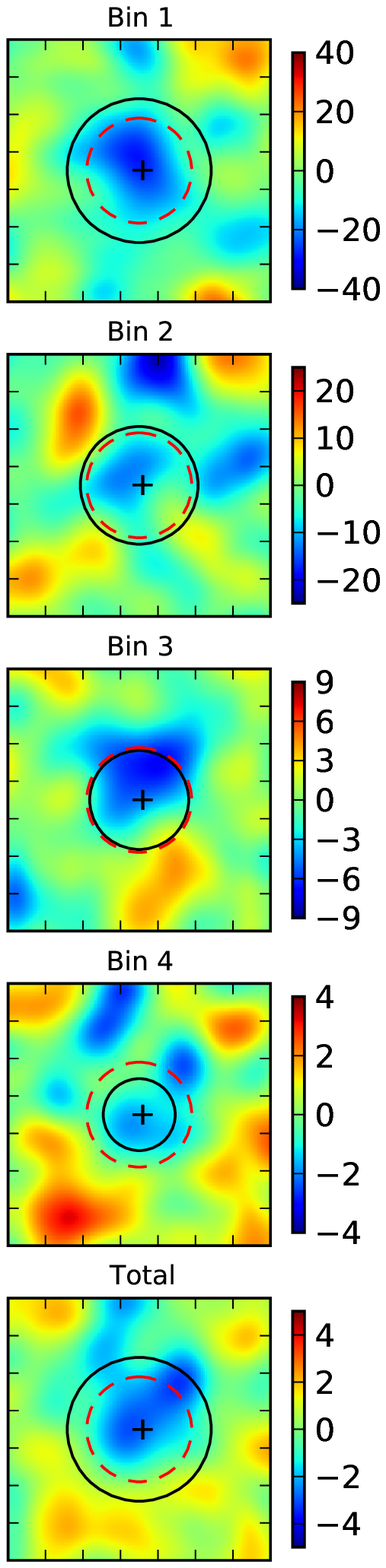}} 
	\subfigure[Radio-quiet LRG Sample]{\includegraphics[scale=0.75]{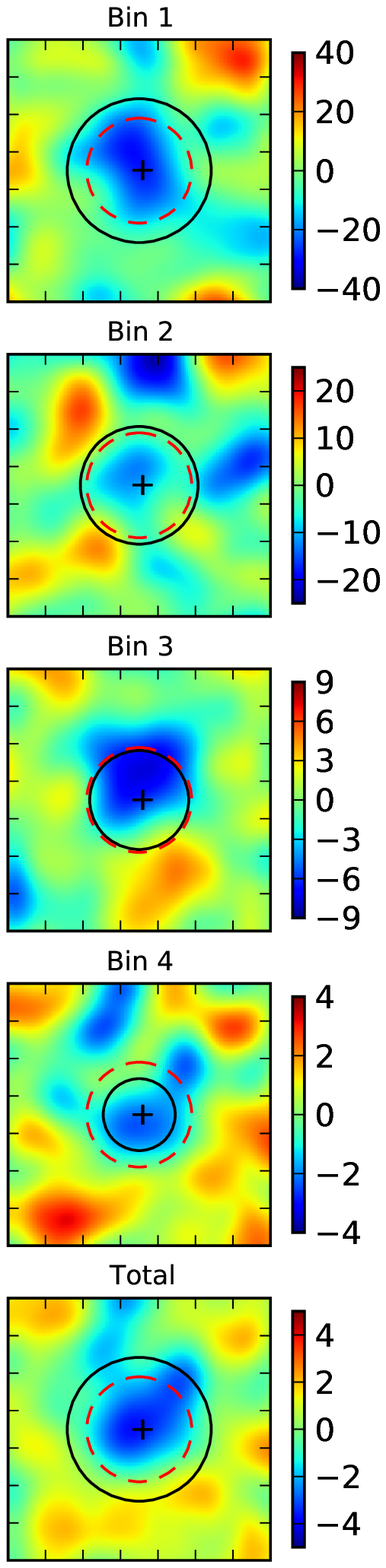}}
	\subfigure[Random Sample]{\includegraphics[scale=0.75]{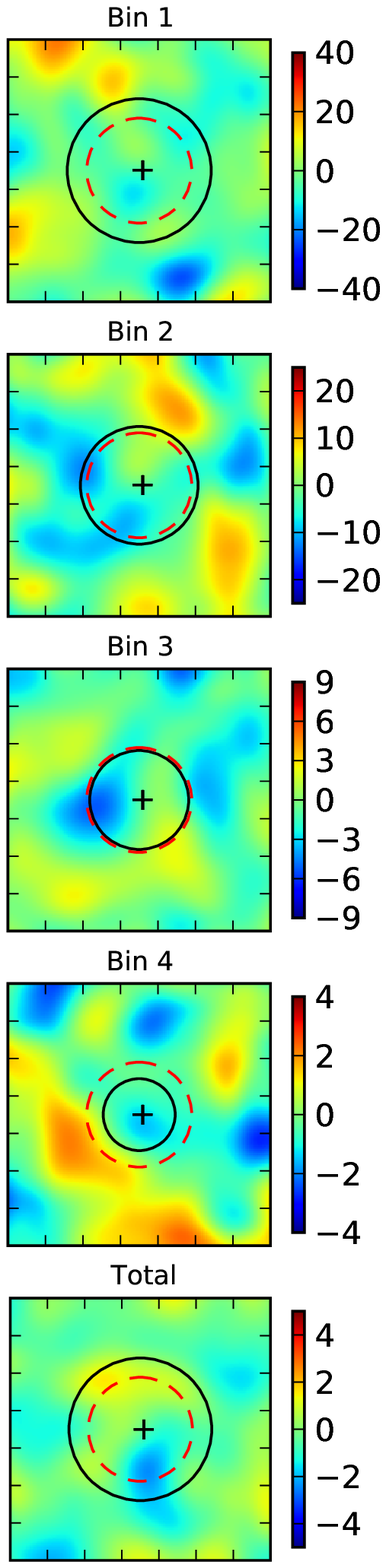}}
	\end{center}
\caption{\label{stacked_images}The stacked filtered temperature maps for each luminosity bin for the (a) full LRG sample, (b) radio-quiet LRG sample and (c) a randomly selected sample. 
Each map is $7' \times 7'$ and is in CMB temperature units $(\micro$K). The values for $\delta T_{\circ}$  and the number of maps in each stack are given in Tables \ref{temp_full_table} and 
\ref{temp_rq_table} for (a) and (b), respectively. The radio-quiet sample excludes the 10\% of the full LRG sample that has been defined as radio-loud (see Section \ref{decrement_results}).
 The stacked random maps serve as a null test of the data, as the SZ signal should be consistent with zero for these maps. The number of maps in each stack for the random sample is equal
  to that of the full sample. The solid circles in each submap mark the characteristic size of a cluster for a given luminosity/mass bin, in terms of $\theta_{200}$. The values for $\theta_{200}$ have been 
  determined from simulation. The dashed circles show the FWHM of the ACT beam function. The crosses mark the center of each map. See online article for color version. }
\end{figure*}

\begin{center}
\begin{figure*}
\includegraphics[scale=0.9]{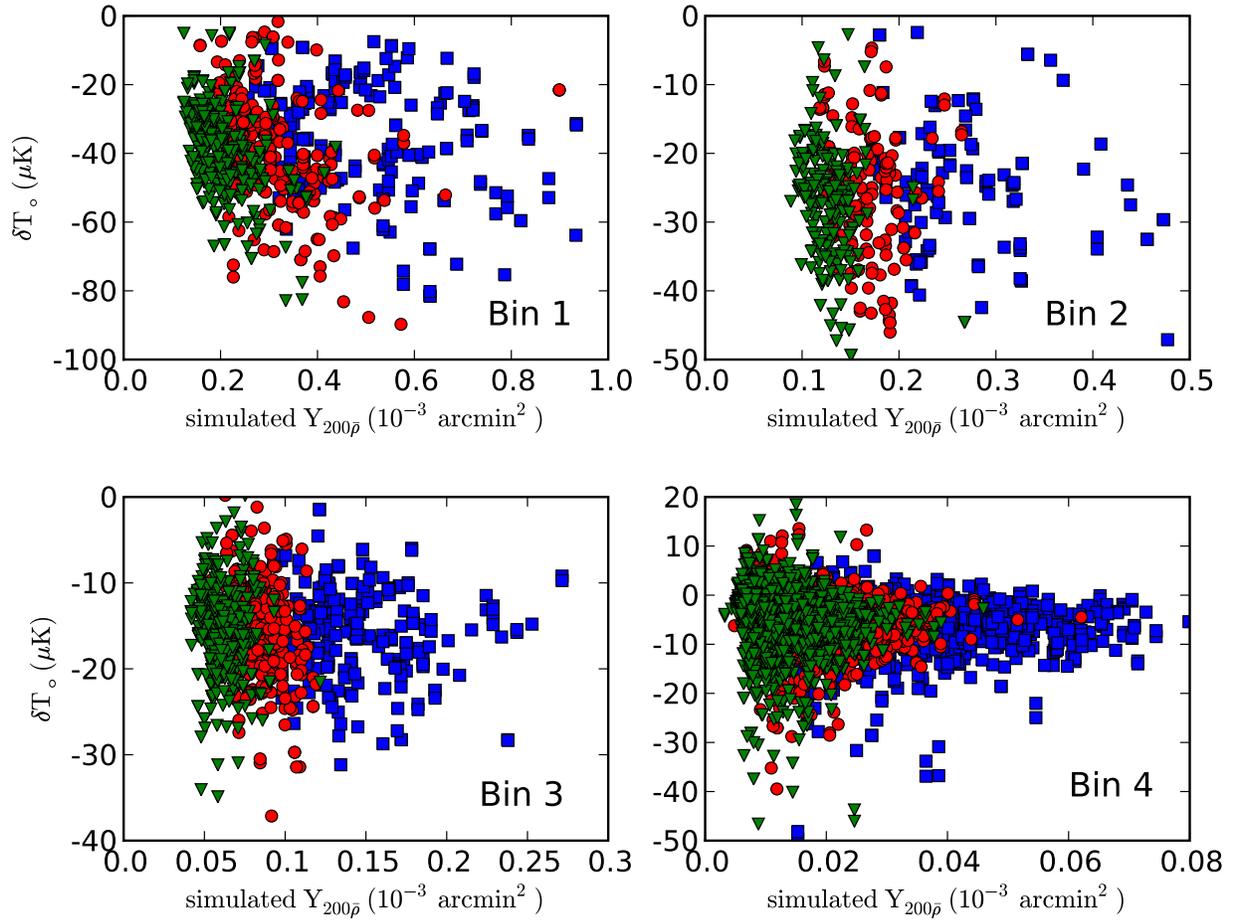}
\caption{\label{redshift_dependence}The redshift dependence of the relation between $\delta T_{\circ}$ and $Y_{200\bar{\rho}}$ for each luminosity/mass bin, as determined from simulations 
described in \citet{sehgal/etal:2009}. The redshift ranges are represented as follows: $z < 0.26$ (squares), $0.26 < z < 0.37$ (circles), and $z >  0.37$ (triangles). The four subplots 
correspond to the luminosity bins used in the LRG stacking analysis.  The mass range for each bin corresponds to the mass estimates determined from analysis of halo bias (Table \ref{mass_table}). }
\end{figure*}
\end{center}

We use numerical simulations \citep{sehgal/etal:2009} to relate the matched filter decrement $\delta T_{\circ}$ to the integrated SZ signal $Y$ (equation \ref{Y_def}). In an unfiltered map, where the 
cluster is larger than the beam, $Y \simeq \delta T_0 \theta_c^2$, where $\theta_c$ is the angular size of the cluster.  For a cluster of fixed physical size, $Y \propto \delta T_0 d_A^{-2}$ 
where $d_A$ is the angular diameter distance to the cluster. Because the filter acts as a $\nabla^4$ operator and removes large-scale signals in the map, the expected relationship for
 nearby clusters of fixed size is likely even steeper. However, rather than rely on this approximate scaling to weight the observations, we apply the same filter used on the ACT data
to the simulated maps and compute $\delta T_{\circ}$ for each of the clusters. The simulated maps are first convolved with the ACT 148 GHz beam function. Then, motivated by the angular diameter 
distance scaling, we fit a model for the $Y-\delta T_{\circ}$ relation which allows for a power-law dependence on redshift,

\begin{equation}
\label{model}
Y_{est, i} = C \  \delta T_{\circ, i} \  z_i^{-\beta},
\end{equation}
where $z_i$ is the cluster redshift and $Y_{est, i}$ serves as an estimate for the SZ signal within a disk of 
radius $R_{200}$ for the $i^{\mathrm{th}}$ cluster. We define $R_{200}$ as the radius within which the average density is equal to 200 times the mean matter density of the 
universe, $\bar{\rho}(z) = \rho_{crit}(z = 0)\Omega_m(1+z)^3$. The redshift dependence in equation \ref{model} for the $i^{\mathrm{th}}$ cluster is determined mainly from the angular size of the cluster profile, 
after convolution with the matched filter.  For a fixed value of $\delta T_{\circ}$, lower redshift clusters appear larger on the sky and have larger values of $Y$
than clusters at higher redshift, as shown in Figure \ref{redshift_dependence}. As we are considering clusters of comparable size to the ACT beam function (see Figure \ref{stacked_images}), 
we expect the integrated SZ signal $Y$ to depend linearly on the matched filter decrement $\delta T_{\circ}$. 

\begin{center}
\begin{figure*}
\includegraphics[scale=0.9]{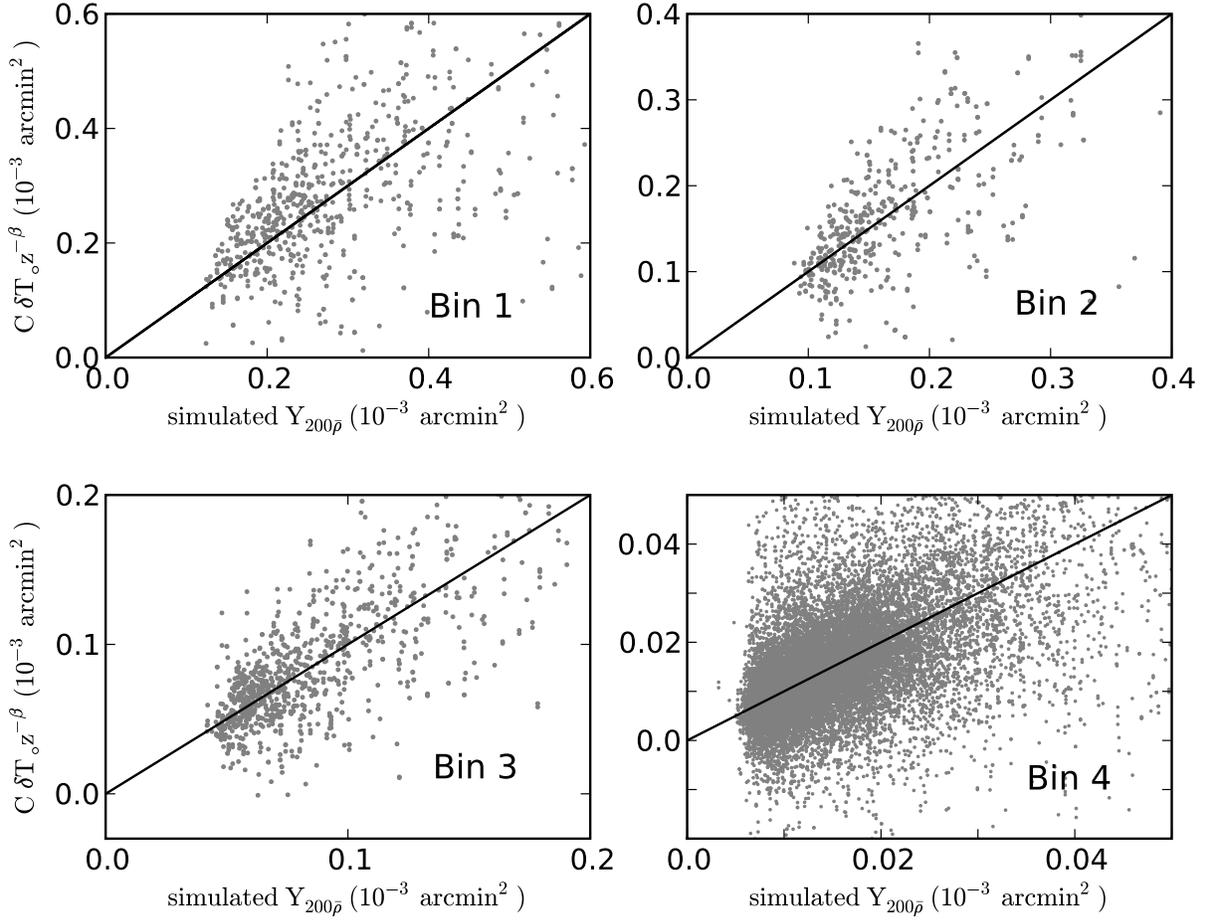}
\caption{\label{Yest_vs_Ysim}The correlation between $\delta T_{\circ}$ and $Y_{200\bar{\rho}}$ for each luminosity/mass bin, as determined from simulations described in \citet{sehgal/etal:2009}. 
The simulation values for $Y_{200\bar{\rho}}$ of the $i^{\mathrm{th}}$ cluster are fit to the model $Y_{est, i} = C \ \delta T_{\circ, i} \ z_i^{-\beta}$. The four subplots correspond to the luminosity bins used 
in the LRG stacking analysis.  The mass range for each bin corresponds to the mass estimates determined from analysis of halo bias (Table \ref{mass_table}). }
\end{figure*}
\end{center}

 The relationship between $Y$ and $\delta T_{\circ}$ is obtained for the mass range of each luminosity bin. Here the
 mass range is represented by the $1\sigma$ uncertainty in the central $M_{200\bar{\rho}}$ value, listed in Table \ref{mass_table}. Due to systematic uncertainties in determining halo
  mass from LRG luminosity, we perform this analysis for two separate groups of mass bins (see Section \ref{mass_estimation}). The best fit values and corresponding errors for $C$ and $\beta$ are 
  summarized in Table \ref{param_table}. The uncertainties reported are the formal $1\sigma$ errors computed using Levenberg--Marquardt least squares minimization. The correlation between $Y_{est}$ 
  and $Y_{200\bar{\rho}}$ is shown in Figure \ref{Yest_vs_Ysim}. The scatter in this correlation results from a combination of noise in the simulated maps and the fact that a cluster's profile and the center of SZ 
  signal is in general not that of the matched filter profile. 

The values for $Y_{200\bar{\rho}}$ are found for the $j^{\mathrm{th}}$ luminosity bin by summing over each LRG in the bin as follows, 

\begin{equation}
\label{Y_200}
Y_{200\bar{\rho}, j} = \frac{\sum_i^{N_j} w_i \ C_j \  \delta T_{\circ, i} \ z_i^{-\beta_j} }{\sum_i^{N_j} w_i},
\end{equation}
where  $\delta T_{\circ, i}$ is the matched filter decrement recovered from the filtered ACT map and $N_j$ is the number of LRGs in the $j^{\mathrm{th}}$ luminosity bin. The weight $w_i$
 is related to the number of observations per pixel (see Section \ref{stacking_algorithm}). Values for $Y_{200\bar{\rho}}$ are reported in Tables \ref{temp_full_table} and \ref{temp_rq_table}. 
 These values correspond to the halo masses $M_{200\bar{\rho}}$ derived from considering luminosity-dependent halo bias. The error in $Y_{200\bar{\rho}}$ 
 represents the uncertainty after propagating errors in $\delta T_{\circ}$, $\beta$, and $C$.   These formal errors do not include the systematic uncertainties in our model which assumes
  that the gas profiles in \citet{sehgal/etal:2009} are accurate representations of the profiles for typical clusters.
  
\begin{table}[t]
\caption{Best-Fit Parameters for the $Y_{200\bar{\rho}}$ - $\delta T_{\circ}$ relation, obtained from simulation and fitted using Equation \ref{model}}
\begin{center}
\label{param_table}
\begin{tabular}{ |c c c c | }
\hline
$N_{\mathrm{bin}}$ &  Mass Range &  $C$  & $\beta$ \\
                &    $10^{14}M_{\odot}$ & $\mathrm{arcmin^2}$/K  & \\ \hline
 &  & Halo Bias &  \\ 
\hline
781   &  3.20 -- 5.32   &  1.41 $\pm$ 0.09 & 1.51 $\pm$ 0.05 \\
470 &  2.69 -- 3.09  & 1.48 $\pm$ 0.09 & 1.31 $\pm$ 0.06 \\
1050 & 1.76 -- 2.05 & 1.32 $\pm$ 0.05 & 1.24 $\pm$ 0.04 \\
19003 & 0.56-- 1.04 &  1.19  $\pm$ 0.03 & 1.07 $\pm$ 0.02 \\
\hline
& & Weak Lensing Measurements & \\
\hline
3127   &  1.72 -- 3.46   & 1.37 $\pm$ 0.03 & 1.30 $\pm$ 0.02 \\
590 &  1.62 -- 2.48  & 1.21 $\pm$ 0.07 & 1.24 $\pm$ 0.05 \\
5408 & 1.12 -- 1.74 & 1.16 $\pm$ 0.03 & 1.20 $\pm$ 0.02 \\
11274 & 0.60 -- 1.02 &  1.12  $\pm$ 0.03 & 1.08 $\pm$ 0.02 \\
\hline
 \end{tabular} 
 \tablecomments{The two groups of mass ranges correspond to the separate methods that are used to obtain halo masses for the LRG sample (see Table \ref{mass_table}).}
 \end{center}
 \end{table}

\subsection{Estimates for Halo Mass $M_{200\bar{\rho}}$}
\label{mass_estimation}
We take two different approaches to estimating the characteristic mass associated with halos of a given luminosity.  The first approach is to use measurements of the 
halo bias, defined as the relation between the galaxy power spectrum $P_{gg}(k)$ and the linear matter power spectrum $P_{\mathrm{lin}}(k)$, 

\begin{equation}
P_{gg}(k)  = b(k)^2 P_{\mathrm{lin}}(k). 
\end{equation}
On the large scales probed by the LRGs, the bias $b(k)$ is expected to be nearly scale-independent, but strongly dependent on luminosity.   \citet{tegmark/etal:2004} and \citet{zehavi/etal:2005} 
have fitted an empirical model for galaxy bias relative to the bias of $L_{\star}$ galaxies\footnote[1]{$L_{\star}$ is defined with respect to the \citet{schechter/1976} luminosity function for SDSS 
galaxies \citep[see][]{blanton/etal:2003}}:

\begin{equation}
\label{relative_bias}
\frac{b}{b_{\star}} = 0.85 + 0.15\frac{L}{L_{\star}} + 0.04(M_{\star} - M_{0.1r}), 
\end{equation}
where $M_{\star} - 5 \mathrm{log_{10}}(h) = -20.44$  \citep{blanton/etal:2003}.  For $h=0.71$, $M_{\star} = -21.18$. Both $M_{\star}$ and $L_{\star}$ are in the $^{0.1}r$-band. 
\citet{percival/etal:2008} have shown this model for relative bias to be a reasonable fit for the LRG sample considered in this paper. The value for $b_{\star}$ used in this analysis is
 $b_{\star} = 1.19$ \citep{reid/etal:2010}.  

\citet{tinker/etal:2010} use numerical simulations to derive a relationship between halo mass and bias.  They identify halos in their cosmological simulation using the  spherical 
overdensity algorithm to define the halo mass: 

\begin{equation}
M_{\Delta} = \frac{4}{3}\pi R_{\Delta}^3\bar{\rho}(z)\Delta, 
\end{equation}
where $\bar{\rho}(z)$ is the mean matter density of the universe at redshift $z$. This analysis uses $\Delta = 200$ and denotes the halo mass by $M_{200\bar{\rho}}$. \citet{tinker/etal:2010}
 estimate the relation between halo bias and $M_{200\bar{\rho}}$:

 \begin{equation}
 \label{bias_vs_mass}
 b(\nu) = 1 - A \frac{\nu^B}{\nu^B + \delta_c^B} + 0.183\nu^{1.5} + C\nu^{2.4}, 
 \end{equation}
 where $\nu$ represents the ``peak height'' of the density field and is given by $\nu = \delta_c/ \sigma(M_{200\bar{\rho}})$, $\delta_c$ is the critical density for collapse, and $\sigma$ is the
  linear matter variance on the scale of the halo, i.e., $R = (3M/4\pi\bar{\rho})^{1/3}$. We use $z = 0$ to determine the matter variance $\sigma$ in order to compare $b(\nu)$ to $b/b_{\star}$.  In all calculations, we use $\delta_c = 1.686$. The parameters in equation \ref{bias_vs_mass} are: 
\begin{eqnarray*}
A & = & 1.0 + 0.24y \ \mathrm{exp}[-(4/y)^4], \\
B & =  & 0.44y - 0.88, \\
C & =  & 0.019 + 0.107y + 0.19 \ \mathrm{exp}[-(4/y)^4], 
\end{eqnarray*}
 where $y \equiv \mathrm{log}_{10}(200)$. 
 
 Using equations \ref{relative_bias} and \ref{bias_vs_mass},  we obtain estimates for halo mass for each luminosity bin. These values are given in Table \ref{mass_table}. We determine 
 mass from luminosity for each LRG in a given bin and report $M_{200\bar{\rho}}$ as the mean of these values. The error in $M_{200\bar{\rho}}$ is estimated by the standard deviation 
 of this mean, within each bin. 
 
 Stacked gravitational lensing measurements are an alternative method of estimating the halo mass-galaxy luminosity relation. \citet{reyes/etal:2008} stack a sample of clusters from the 
 SDSS maxBCG catalog \citep{koester/etal:2007a, koester/etal:2007b} and use weak gravitational lensing measurements to investigate the relation between cluster mass and various optical 
 tracers. Since LRGs trace massive halos \citep{ho/etal:2009},  the maxBCG catalog should represent a population similar to the LRG sample used in this work. Noting that \citet{reyes/etal:2008}
  bin their cluster sample by $^{0.25}r$-band BCG luminosity, we compute the $^{0.25}r$-band luminosity for each LRG and determine halo masses using equation 15(c) of \citet{reyes/etal:2008}.  
  Table \ref{mass_table} lists these mass values, as well as those obtained from analysis of halo bias. The range in these two mass estimates represents some of the systematic uncertainties in relating LRG 
  luminosity to halo mass.

\section{SZ--Mass Scaling Relation}
\label{scaling_relation}
 
\begin{center}
\begin{figure*}
\includegraphics[scale=0.9]{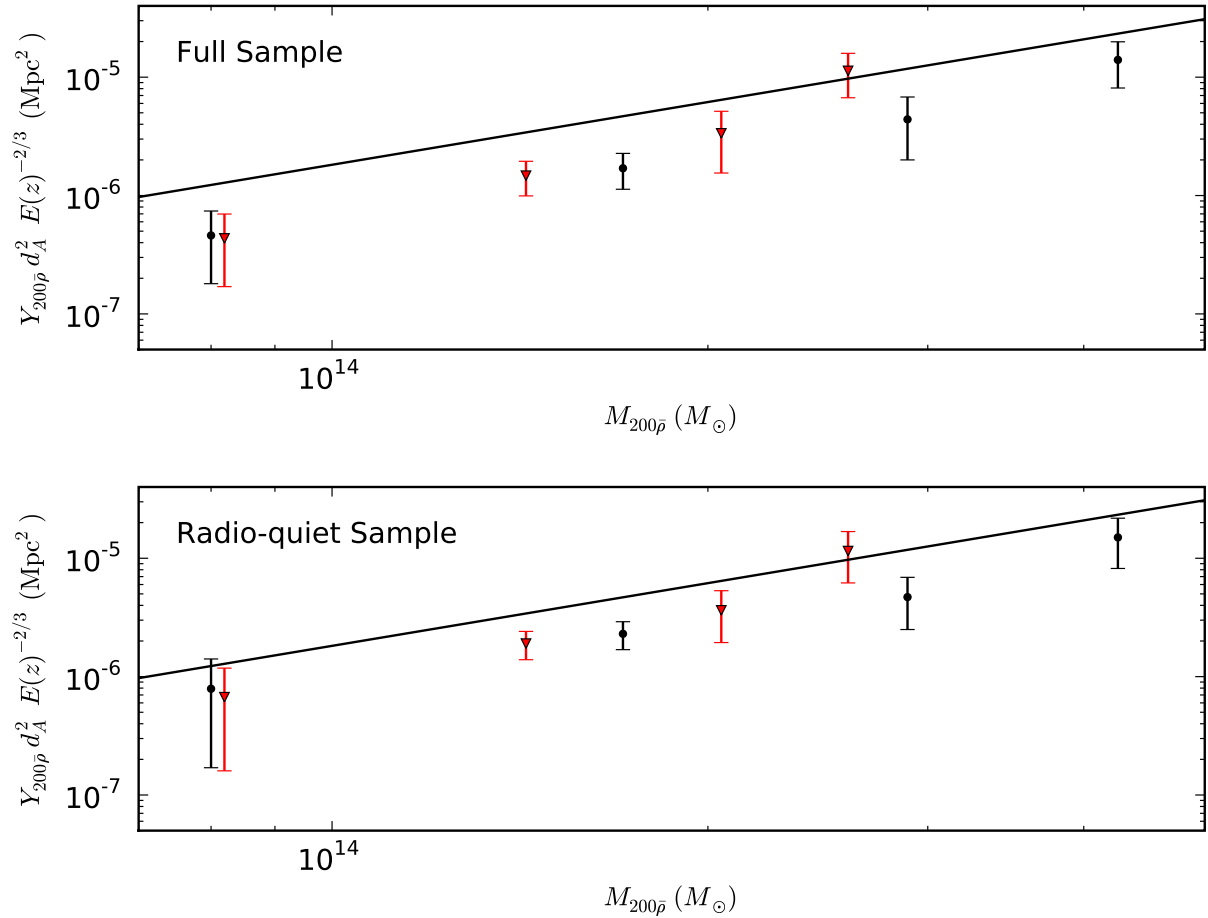}
\caption{\label{Y_vs_M}  SZ -- Mass Correlation. $Y_{200\bar{\rho}}$ estimates are plotted against cluster masses for each luminosity bin for both the full (Table \ref{temp_full_table}) and 
radio-quiet  (Table \ref{temp_rq_table}) LRG samples  . The two sets of points shown use separate mass estimates for the four LRG luminosity bins. The black circles correspond 
to mass estimates derived from analysis of halo bias. The triangles (red in online version) correspond to mass estimates derived from weak gravitational lensing measurements
  \protect\citep{reyes/etal:2008}. The range in these masses represents some of the systematic uncertainties in converting LRG luminosity to halo mass. The solid line shows the expected
   model for the $Y-M$ relation (equation \ref{sim_relation}), as determined from the microwave  sky simulations of \citet{sehgal/etal:2009}. }
\end{figure*}
\end{center}

In this section, we present an analysis of the correlation between cluster mass and integrated SZ signal.  Specifically, we compare $Y_{200\bar{\rho}}$ to the cluster mass $M_{200\bar{\rho}}$
 for the luminosity/mass bins in Table \ref{mass_table}. For self-similar evolution, analytical theory \citep[e.g.,][]{komatsu/seljak:2001} predicts a simple relationship between $Y$ and cluster mass. 
 Assuming virial and hydrostatic equilibrium, the cluster gas temperature can be related to $M$ and $E(z)$ through \citep[e.g.,][]{bryan/norman:1998}
 
\begin{equation}
T \propto M^{2/3} E(z)^{2/3}, 
\end{equation}
where for a flat $\Lambda$CDM cosmology

\begin{equation}
E(z) = [\Omega_m (1 + z)^3 + \Omega_{\Lambda}]^{1/2}.
\end{equation}
Under assumptions of an isothermal ICM, 

\begin{equation} 
Y_{200\bar{\rho}} \propto d^{-2}_A(z) \ M_{200\bar{\rho}} \  T , 
\end{equation}
where $d_A(z)$ is the angular diameter distance to the cluster. Combining the above equations leads to the expectation for self-similar evolution

\begin{equation}
Y_{200\bar{\rho}} \propto  d_A^{-2} \ M_{200\bar{\rho}}^{5/3}  \ E(z)^{2/3}.
\end{equation}
Anticipating this scaling relation, we have plotted $Y_{200\bar{\rho}}d_A^2E^{-2/3}$ against the halo masses $M_{200\bar{\rho}}$ in Figure \ref{Y_vs_M}.  We show the two separate halo 
mass estimates for a given bin to illustrate the uncertainties that exist in converting LRG luminosity to halo mass. The results for both the full (Table \ref{temp_full_table}) and radio-quiet (Table 
\ref{temp_rq_table}) LRG samples are shown in this figure. The expected $Y-M$ relation as determined from numerical simulations is shown as a solid line and is given by:

\begin{equation}
\label{sim_relation}
Y_{200\bar{\rho}} d_A^2E^{-2/3} = 10^{-\gamma} \left ( \frac{M_{200\bar{\rho}}}{10^{14} M_{\odot}} \right )^{\alpha}, 
\end{equation}
where the best fit parameters are $\alpha$ = 1.76 and $\gamma$ = 5.74. These parameters are obtained from the microwave sky simulations described in \citet{sehgal/etal:2009}, where halos of 
mass $M_{200\bar{\rho}} > 2.82 \times 10^{13}  \ M_{\odot}$ have been considered. Furthermore, the redshift range of the clusters is restricted to $0.15 < z < 0.50$, and a cylinder definition of $Y$
 is used. Tables \ref{temp_full_table} and \ref{temp_rq_table} give the values of $Y_{200\bar{\rho}}d_A^2E^{-2/3}$ and associated errors for the full and radio-quiet samples. These estimates are 
 determined using a summation similar to equation \ref{Y_200}, where the redshift of each LRG is used to calculate $d_A(z)$ and $E(z)$. They correspond to the masses derived from considering 
 luminosity-dependent halo bias. 
\newline

 \section{Discussion}
 \label{conclusion}
 
By stacking LRGs, we have obtained estimates of the SZ signal as a function of mass for groups and clusters. The stacking analysis enabled the detection of an SZ signal for clusters with
 masses around $10^{14} M_\odot$, and the amplitude of the signal is consistent with our theoretical estimates. Using masses derived from halo bias, the slope of the 
 inferred $Y - M$ relation is consistent with expected results.  However, when using stacked weak gravitational lensing measurements to estimate the halo masses, the slope appears to be slightly 
 steeper than the self-similar prediction. 
 
 For both mass calibrations, the overall normalization of the $Y-M$ relation appears lower than expected.\footnote[1]{Note that immediately following the submission of this work, the {\em{Planck}} collaboration reported results on SZ scaling relations for both optical- and X-ray-selected galaxy clusters \citep{planck:2011b, planck:2011a}. Its analysis of the optical scaling relations finds a similar deficit in the overall normalization, while the scaling relations for X-ray clusters are in agreement with expected results.}  As the SZ effect traces the thermalized cluster pressure, there exists the possibility that there are significant non-thermal pressure components supporting the halo mass. However, the normalization of the $Y-M$ relation depends upon the method of obtaining halo mass from 
 LRG luminosity, which is subject to large uncertainties. Applying an analysis of halo occupation distribution to the SDSS LRG sample, \citet{zheng/etal:2009} find a $\sim$16\% log-normal scatter in the central LRG luminosity for halos of a given mass. This scatter applies only to central LRGs and does not include the $\sim$5\% of galaxies that are satellites. Thus, the scatter in the LRG luminosity for a fixed halo mass is likely even larger.   Furthermore, there exists uncertainties in the mass estimation from weak lensing measurements \citep[e.g.,][]{mandelbaum/etal:2008, rozo/etal:2009} that contribute to the discrepancy in our mass calibrations.
 
We have detected radio source contamination of the SZ signal at 148 GHz for $\sim$10\% of the LRGs. This suggests the possibility that infrared sources may also have a contamination
 effect on our measurement of the SZ signal for the LRG sample.  \citet{mittal/etal:2009} found that the likelihood for a cluster to host a radio-loud BCG is a strong function of the central cooling 
 time, in the sense that all strong cool-core clusters contain a radio-loud BCG. Thus, our radio-quiet LRG sample may select against clusters/groups with a strong cool core. However, in his 
 analysis of a large sample of clusters and groups at $z \le 0.1$, \citet{sun:2009} did not find any radio-loud central galaxies in groups with strong cool cores.  If this is true at higher redshift, our
  radio-quiet LRG sample would be representative of all groups and clusters.
 
Another source of systematic error is the conversion of the measured decrement $\delta T_{\circ}$ to the integrated SZ signal $Y$.  Our approach relies on simulations from
 \citet{sehgal/etal:2009} to estimate the effect of the matched filter on smoothing the cluster signal. Future work will explore this uncertainty by using simulations with different models of cluster
  gas physics. 
 
This work is a first application of a novel stacking method for searching for the SZ effect in clusters by applying a full optical catalog to a CMB survey. Both data sets are rich and have multiple 
components, with possible correlations between them, that we will be sorting out as the data and our understanding improves. Future ACT analysis of the SZ effect from galaxy clusters will 
include data from multiple seasons and millimeter bands and will include the new SDSS DR8 analysis.  

\begin{acknowledgments}
This paper is part of a senior thesis supervised by D. Spergel. ACT operates in the Parque Astron\'{o}mico Atacama in northern Chile under the auspices of Programa de Astronom\'{i}a, a program of the Comisi\'{o}n Nacional de Investigaci\'{o}n Cient\'{i}fica y Tecnol\'{o}gica de Chile (CONICYT).  We thank Masao Uehara for coordinating ACT's operations in Chile and  Paula Aguirre, Bill Page, David Sanchez and Omean 
Stryzak for assistance in operating the telescope.  We would also like to thank Eyal Kazin, Will Percival, Beth Reid and Reina Reyes for useful discussions during the development of this work. 

This work was supported by the U.S. National Science Foundation through awards AST-0408698 for the ACT project, and PHY-0355328, AST-0707731 and PIRE-0507768. Funding was also 
provided by Princeton University and the University of Pennsylvania. JD acknowledges support from a RCUK Fellowship. Computations were performed on the GPC supercomputer at the 
SciNet HPC Consortium. SciNet is funded by: the Canada Foundation for Innovation under the auspices of Compute Canada; the Government of Ontario; Ontario Research Fund - Research 
Excellence; and the University of Toronto.  SD, AH, and TM were supported through NASA grant NNX08AH30G. ADH received additional support from a Natural Science and Engineering 
Research Council of Canada (NSERC) PGSD scholarship. AK was partially supported through NSF AST-0546035 and AST-0606975, respectively, for work on ACT. ES acknowledges 
support by NSF Physics Frontier Center grant PHY-0114422 to the Kavli Institute of Cosmological Physics. NS is supported by the U.S. Department of Energy contract to SLAC no. 
DE-AC3-76SF00515. SD acknowledges support from the Berkeley Center for Cosmological Physics.
\end{acknowledgments}

\bibliography{act_lrg}
\end{document}